# Conception d'outils de communication spécifiques au contexte éducatif

▶ Sébastien GEORGE (LIESP, INSA-Lyon),
Cécile BOTHOREL (TECH/EASY, France Telecom R&D Lannion)

■ **RÉSUMÉ** • Dans un contexte de formation à distance, le simple fait de fournir des outils classiques de communication (forum, chat, …) n'est pas toujours suffisant pour faire émerger des interactions entre apprenants et ainsi favoriser la construction collective de connaissances. Une solution consiste alors à proposer des activités collectives qui incitent les apprenants à communiquer. Mais, même dans ce cas, les outils peuvent parfois devenir un frein à la communication. Nous présentons dans cet article des expériences de conception d'outils de communication particuliers qui ont pour but de faciliter ou d'orienter les communications dans un contexte éducatif, voire même d'inciter les interactions en dehors de toute activité collective prescrite. Nous décrivons des outils de communication synchrone (*chats* semi-structurés) et asynchrone (forum temporellement structuré, forum contextuel), ainsi qu'un système facilitant l'entraide entre apprenants d'une formation en ligne.

■ **MOTS CLÉS** • conversation synchrone et asynchrone, chat, forum, formation en ligne, structuration des discussions, réseau d'entraide.

■ **ABSTRACT** • *In a distance learning context, providing usual communication tools (forum, chat, …) is not always enough to create efficient interactions between learners and to favour collective knowledge building. A solution consists in setting-up collective activities which encourage learners to communicate. But, even in that case, tools can sometimes become a barrier to communication. We present in this paper examples of specific tools that are designed in order to favour and to guide communications in an educational context, but also to foster interactions during learning activities that are not inherently collaborative. We describe synchronous communication tools (semi-structured chat), asynchronous tools (temporally structured forum, contextual forum) and a system which promotes mutual aid between learners.*

■ **KEYWORDS** • *synchronous and asynchronous discussion, chat, forum, online learning, discussion structuring, mutual aid network.*

## 1. Introduction

L'émergence et le développement des apprentissages à distance entraînent de nouveaux enjeux à la fois éducatifs et sociaux. À travers les plates-formes de formation à distance, les acteurs (apprenants, tuteurs, concepteurs, administrateurs) partagent des ressources (contenus de formations), des outils de communication (courrier électronique, forum, chat, etc.) et éventuellement d'autres services (agenda, outils de travail collaboratif).

Les expériences de dispositifs informatiques fournis aux apprenants montrent bien souvent que ces derniers ne parviennent pas à en tirer tout le bénéfice escompté par les concepteurs ou les enseignants. En particulier, les possibilités d'interactions sont la plupart du temps surestimées. Les conversations qui sont médiatisées ne se créent pas de manière aussi spontanée qu'en présence. Les communications et interactions réellement collaboratives et riches d'un point de vue apprentissage ne sont pas si fréquentes (Gommer et Visser, 2001 ; Hotte et Pierre, 2002). Dans cet article, partant de l'hypothèse qu'une partie des difficultés peut provenir des outils eux-mêmes, nous discutons de recherches visant la conception d'outils d'interactions spécifiques offrant des fonctionnalités adaptées à la problématique de l'enseignement ou de la formation à distance.

## 2. Les problématiques de recherche en lien avec les interactions pédagogiques en ligne

D'une manière générale, nous considérons que les recherches dans le domaine des interactions pédagogiques en ligne ont des spécificités par rapport aux recherches dans le domaine de la CMO (Communication Médiatisée par Ordinateur ou son équivalent anglais CMC-Computer-Mediated Communication). Comme signalé par Dejean-Thircuir et Mangenot (2006), « des questions spécifiques, d'ordre éducatif, viennent complexifier le tableau, celle des outils développés à des fins pédagogiques (…), celle des tâches données à réaliser aux apprenants, celle enfin des modalités de réalisation de ces tâches, pouvant aller de la simple mutualisation à la collaboration par petits groupes ». Bien entendu, il est possible de s'appuyer sur les études et les résultats du domaine de la CMO mais les recherches menées dans le domaine des interactions pédagogiques en ligne ont aussi leurs caractéristiques propres.

Les interactions pédagogiques en ligne constituent un objet d'étude auxquels s'intéressent plusieurs disciplines. A l'instar de Mangenot (2006), lui-même s'appuyant sur Gagné *et al.* (1989), nous pouvons distinguer quatre grands types de recherche concernant les interactions pédagogiques en ligne :
- Type 1 : théoriser,
- Type 2 : observer/décrire,
- Type 3 : expliquer/prouver,
- Type 4 : transformer/développer des outils de communication.

Les quatre types ne sont pas cloisonnés. Par exemple, une recherche peut tout à fait tenter de théoriser un phénomène (type 1) puis tenter de l'expliquer avec une expérience (type 3). Cependant une recherche aura le plus souvent un type dominant.

La théorisation des situations d'interactions en ligne consiste par exemple à définir/étudier des modèles d'interactions (Peraya et Ott, 2001) ou à expliciter des notions particulières (dispositif, rôles, distance, etc.). La plupart du temps, ce type de recherche théorique ne s'appuie pas sur des analyses concrètes d'interactions.

Le deuxième type de recherche vise à observer et décrire des nouvelles pratiques en relation avec les outils qui les instrumentent. Les recherches de ce type adoptent le plus souvent une démarche qualitative. Les principales disciplines concernées sont les sciences du langage, les didactiques, la sociologie et les sciences de l'information et de la communication. L'objectif est de mener des études à partir de situations écologiques, c'est-à-dire non réalisées pour le besoin d'une expérimentation. Par exemple, un certain nombre de travaux tentent d'observer et de décrire les impacts des dispositifs techniques sur la communication dans un contexte éducatif (Herring, 2004 ; Dejean-Thircuir et Mangenot, 2006). D'autres recherches tentent d'analyser les forums électroniques pour identifier les constructions collaboratives de connaissances (Desjardins, 2002).

Le troisième type fait référence à des recherches que nous pouvons qualifier d'expérimentales. L'objectif principal de ce type de recherche est d'obtenir des résultats mesurables à partir de traitements quantitatifs (contrôle de variables, catégorisation, etc.). Le but final est d'expliquer un phénomène ou de prouver une théorie. La plupart du temps, les données traitées ne sont pas issues d'un milieu écologique mais proviennent d'un dispositif expérimental. Par exemple, la mise en place d'une expérimentation visant l'étude des représentations des connaissances dans un forum, est décrite dans (Gettliffe-Grant, 2003).

Enfin, le quatrième type de recherche, celui qui retient notre attention dans cet article, concerne la conception et le développement d'outils de communication spécifiques au contexte éducatif. Ce type regroupe des recherches qui peuvent être considérées comme des recherches-actions ou des recherches-développements qui s'appuient en général sur des conclusions ou recommandations issues des types précédents. Les travaux qui rentrent dans cette dernière catégorie concernent principalement des chercheurs en informatique dont l'objectif principal est de modéliser et de concevoir des systèmes de communication particuliers. Nous pouvons répertorier quatre principaux axes dans ce type de recherche :

– faciliter le déroulement des interactions : par exemple des outils avec des présentations alternatives des fils de discussion,

– orienter les interactions vers des formes particulières, ce qui revient à contraindre pour atteindre des objectifs pédagogiques spécifiques (par exemple : apprendre à argumenter),

– inciter la création d'interactions, c'est-à-dire chercher à provoquer des mises en relation entre acteurs,

– faciliter l'analyse automatique des interactions, les outils de communication étant alors développés pour permettre un type d'analyse particulier.

Les outils issus de ces recherches ne sont pas neutres, si tant est qu'il y en ait de neutre, dans les discussions car l'objectif mis en avant est d'influencer les interactions dans une finalité pédagogique. Même les outils provenant du quatrième axe influent sur les interactions car l'analyse des discussions peut par exemple servir à de l'assistance cognitive ou metacognitive (Jermann *et al.*, 2001 ; Dimitracopoulou et Bruillard, 2007).

Dans cet article, nous nous intéressons aux recherches visant le développement d'outils de communication spécifiques en nous focalisant particulièrement sur les structurations particulières proposées dans les outils (structuration des messages les uns par rapport aux autres, structuration des visualisations, etc.). Les travaux que nous présentons dans cet article illustrent les principales approches de ce type de recherches. Comme ces recherches sur la conception d'outils couvrent le plus souvent plusieurs des quatre axes présentés ci-dessus, nous avons choisi comme plan de présentation un autre niveau : la temporalité des échanges. En effet, les discussions synchrones et asynchrones n'ont pas les mêmes contraintes et les structurations proposées s'en trouvent bien différentes. Ainsi, dans la partie 3 de cet article, nous décrivons deux outils de communication synchrone utilisant une interface structurée par des actes de langage. La partie 4 traite des forums de discussions asynchrones et pré-sente deux outils structurant les discussions, d'une part, de manière temporelle et, d'autre part, en fonction du contexte. Enfin, la partie 5 va au-delà des aspects communications en se préoccupant des outils de type *social networking*. Cette partie est notamment illustrée par un exemple de système visant à favoriser l'émergence de réseaux d'entraide.

Dans cet article, nous limitons notre étude aux outils informatiques pour les conversations écrites seulement et non pour les communications orales ou audiovisuelles. Les raisons de ce choix sont multiples. Tout d'abord, l'écrit est encore le mode de communication le plus utilisé dans le contexte de la formation en ligne. Même si les configurations matérielles et les connexions réseaux permettent désormais des communications audiovisuelles, l'écrit possède des avantages dans un contexte éducatif. Ainsi, une étude menée pour comparer les modes de communication écrite et orale lors de collaborations à distance synchrone entre étudiants (Vera *et al.*, 1998) montre que des étudiants utilisant uniquement la communication écrite ont des discussions plus « réfléchies » que ceux qui utilisaient la vidéoconférence. Dans le même sens, il a été montré que la communication écrite permet davantage de réflexion de la part des apprenants en impliquant un processus cognitif plus profond : « *one advantage of text-based communication is that written communication tends to be more reflective than spoken interaction* » (Sherry, 1998). Enfin, l'écrit est un moyen bien adapté aux apprenants timides, pensifs ou hésitants (Berge, 1997).

## 3. Conception d'outils de communication synchrone structurée

Cette partie concerne la conception d'outils informatiques qui permettent l'échange instantané de messages entre plusieurs personnes engagées dans une activité éducative. Ce moyen de communication est caractérisé par le fait que les messages s'affichent en temps réel. Nous commençons cette partie en mettant en évidence des problèmes récurrents survenant avec des outils standards.

### 3.1. La structuration des communications pour améliorer les *chats*

Un grand nombre d'outils de communication synchrone existent sur le marché. Ces outils sont plus communément nommés « *chat* ». Cependant, ces outils de chat n'ont pas beaucoup évolué depuis une vingtaine d'années et demeurent peu adaptés pour des conversations soutenues. Leur fonctionnement repose sur un empilement des messages des utilisateurs de façon temporelle. Nous pouvons formuler des critiques à ce fonctionnement. Tout d'abord, cet empilement de messages les uns à la suite des autres selon leur ordre d'arrivée pose deux problèmes majeurs dans un contexte de conversation synchrone. Le premier problème provient du temps nécessaire à la composition des textes (temps de frappe) qui ne permet pas de répondre de façon immédiate à un message. De ce fait, les temps de latence provoquent des imbrications des interventions qui rendent le suivi de la discussion difficile. Le second problème se situe ainsi dans la lecture des chats, deux messages pouvant se retrouver proches à l'interface alors qu'ils ne sont pas forcément liés et, à l'inverse, deux messages en relation peuvent être séparés par d'autres messages. Ce phénomène s'accentue quand le nombre de personnes participant à la discussion augmente.

La conséquence de cette difficulté de lecture dans les chats classiques est que la conversation est délicate à suivre et à mener. Des résultats d'analyse de discussions médiatisées par des outils de chats ont montré que celles-ci étaient très souvent incohérentes et qu'on trouvait de nombreux messages de « réparation » de la discussion (Herring, 1999). Pourtant, les chats sont des outils très populaires. En fait, les utilisateurs tentent de s'adapter au médium par exemple en abrégeant au maximum les temps de composition des messages (au détriment de l'orthographe bien souvent) ou en se définissant des signes pour coordonner les tours de paroles (Herring, 1999).

Nous présentons dans cette partie deux travaux de recherche ayant abouti à la conception d'outils de communication synchrone visant entre autres à réduire le problème décrit ci-dessus à l'aide d'une structuration des messages. Comme nous le montrons par la suite, la structuration peut également avoir une vocation pédagogique en incitant des types particuliers d'interactions. Ainsi, en repartant de la classification présentée dans la partie 2, les deux outils présentés couvrent trois des quatre axes : d'une part « orienter vers des formes d'interactions particulières » pour l'outil C-chene (partie 3.3) et, d'autre part « faciliter le déroulement des interactions » et « faciliter l'analyse automatique des interactions » pour l'outil SPLACH (partie 3.4).

Dans les deux cas qui vont être présentés, la structuration repose sur la définition d'actes de langage, concept que nous décrivons ici de manière synthétique.

### 3.2. Discussion et actes de langage

L'étude de la fonction du langage a fait l'objet de nombreuses recherches. Pendant longtemps, le langage a été perçu comme étant un moyen de rendre compte et de décrire le monde. Austin (Austin, 1962) présente une autre vision du langage en considérant les énoncés d'une conversation comme étant non plus seulement descriptifs mais comme des actes en soi. Les énoncés sont eux-mêmes des actions et possèdent une force illocutoire, c'est-à-dire une certaine valeur (question, proposition, etc.). Ainsi, une conversation peut être considérée comme une succession d'échanges composés d'interventions de locuteurs, constituées elles-mêmes d'actes de langage dont l'un est l'acte directeur qui détermine la force illocutoire de l'intervention (Roulet *et al.*, 1985).

Il est toutefois très difficile d'analyser le contenu des interventions d'une discussion pour tenter de déterminer ces actes de langage. Malgré les progrès effectués dans le domaine du traitement automatique du langage naturel, les résultats ne sont pas toujours très fiables. Afin de déterminer l'acte de langage directeur dans une intervention, une solution consiste à le demander à l'utilisateur. C'est un principe de base qui est notamment utilisé dans le domaine des EIAH : si le système ne sait reconnaître l'information nécessaire pour son analyse alors il faut la demander à l'utilisateur : « *avoid guessing, get the student to tell you what you need to know* » (Self, 1988). Cette solution amène alors à utiliser une interface structurée pour demander un acte de langage à l'utilisateur avant que celui-ci ne saisisse le contenu de son message de manière libre.

Les deux exemples de systèmes présentés ci-dessous sont fondés sur une interface structurée utilisant des actes de langage. Le premier vise essentiellement à favoriser les interactions entre apprenants et le second intègre également une analyse automatique des conversations.

### 3.3. C-Chene : un outil orientant la communication

Baker et Lund (1997) se sont intéressés aux interactions épistémiques dans les environnements d'apprentissage coopératif. Leurs travaux de recherche les ont amenés à concevoir différents prototypes pour étudier la mobilisation et l'élaboration de notions scientifiques par des élèves.

En particulier, ces chercheurs ont étudié la résolution de problèmes entre élèves dans le domaine de la physique, plus précisément dans la construction de chaînes cinématiques. Dans ce cadre, un environnement support à l'apprentissage coopératif, nommé C-CHENE, a été conçu. Cet environnement comporte un espace de travail, dans lequel les élèves peuvent construire une chaîne énergétique, et un espace de communication (Figure 1). Deux interfaces de communication différentes ont été développées. La première interface permet de communiquer en tapant du texte librement (outil de chat classique) alors que la deuxième interface structure les communications aux moyens d'« actes communicatifs ». L'interface structurée (en bas de la figure 1) comporte vingt-quatre actes communicatifs divisés en quatre sections : faire la chaîne cinématique (« Je propose de … », « Pourquoi ? », etc.), se mettre d'accord (« D'accord », « Es-tu d'accord ? », etc.), gérer l'interaction (« Attends », « Par quoi on commence ? », etc.) et faire autre chose (« Regarde l'expérience », « Lis la feuille » ).

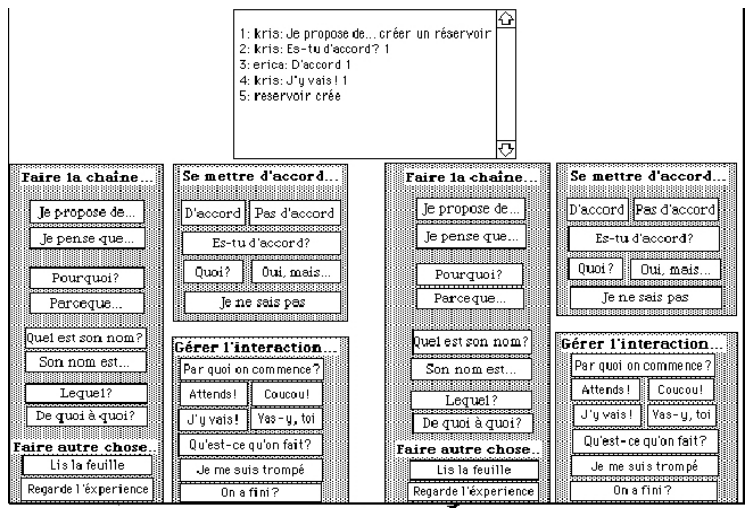

**Figure 1 • L'interface de communication structurée dans C-Chene**

Des expérimentations ont été effectuées en laboratoire avec des binômes d'élèves de 1$^{\text{ère}}$ scientifique pour comparer l'utilisation de l'interface contenant des actes communicatifs à celle d'un chat classique. Il est ressorti de cette expérimentation que l'interface structurée facilitait et encourageait des interactions plus exploratoires et davantage centrées sur la tâche à réaliser (Baker et Lund, 1997). Le fait de contraindre la communication peut donc avoir des effets positifs : « Un juste degré de contrainte sur la communication médiatisée par ordinateur favorise des interactions plus focalisées sur la réflexion et les concepts fondamentaux en jeu » (Baker *et al.*, 2001).

En observant l'interface de communication proposée, nous pouvons cependant voir que les actes communicatifs proposés sont très liés au domaine (« quel est son nom... ») ou à la tâche (« regarde l'expérience », « vas-y toi »). Cette interface ne peut donc être utilisée de manière générique et nécessite de redéfinir les « ouvreurs de phrase » en fonction d'une activité pédagogique cible.

### 3.4. SPLACH : un outil de conversation structurée analysable

Une recherche sur le support à l'apprentissage par projet à distance (George, 2001) a abouti à la conception d'un environnement informatique spécifique nommé SPLACH (Support d'une pédagogie de Projet pour l'Apprentissage Collectif Humain). Cet environnement intègre des outils nécessaires aux activités de projets collectifs (outils de communication asynchrones et synchrones, outil de partage d'application, outil de planification, outil de documentation). En particulier, l'outil de conversation synchrone de SPLACH a été spécialement conçu pour répondre à deux besoins : faciliter les discussions et permettre l'analyse automatique de celles-ci. Les spécificités tiennent dans une présentation arborescente des messages et dans une interface structurée par des actes de langage.

Les actes de langage intégrés dans l'outil de conversation structuré de SPLACH se répartissent en cinq catégories. Trois de ces catégories se trouvent par exemple dans les travaux de Bilange (1991) à savoir : les actes « initiatifs », les actes « réactifs » et les actes « évaluatifs ». Deux autres catégories ont été ajoutées à ces trois catégories de bases : les actes de « salutation » et les actes « auto-réactifs ». Au final, dix actes de langage se répartissent dans ces cinq catégories. Par ailleurs, l'outil de conversation proposé dans SPLACH tente de respecter les contraintes d'enchaînement qui existent entre les actes de langage. L'outil repose en particulier sur l'idée de « paire adjacente » introduite par Clark et Shaefer (1989). Selon ces auteurs, une paire adjacente se compose de deux expressions ordonnées produites par deux locuteurs, la forme et le contenu de la deuxième partie de la paire étant dépendante de la première partie. Dans SPLACH, les paires adjacentes peuvent être représentées sous forme d'un graphe de successions (Figure 2) qui définit une structure arborescente pour une conversation. Le choix de l'ensemble de ces actes est le résultat d'un processus itératif incluant des expérimentations auprès d'utilisateurs. Le nœud de gauche représente l'initiation d'une nouvelle discussion. Ainsi, un utilisateur peut commencer une discussion par une salutation (« saluer ») ou par un acte initiatif (« demander », « proposer » ou « affirmer »). À partir d'un acte initiatif, un utilisateur peut par exemple « répondre » à une demande, « questionner » une affirmation ou bien encore évaluer une proposition (« approuver » ou « désapprouver »). Lorsqu'un utilisateur a fait une intervention, quel que soit son type, il peut réagir à sa propre intervention (actes auto-réactifs) en « précisant » ou « rectifiant ». Nous pouvons par ailleurs remarquer que le graphe de successions ne présente pas de nœud de fin. Cette caractéristique vise à permettre aux débats de se dérouler. Cet aspect est important dans un contexte d'apprentissage pour favoriser l'approfondissement des idées (ici l'objectif n'est pas que les utilisateurs trouvent un accord le plus rapidement possible mais qu'ils puissent échanger et partager leurs connaissances).

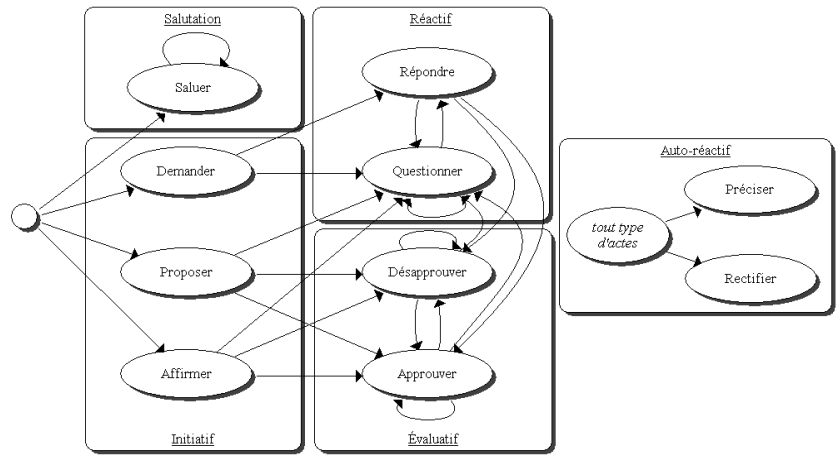

**Figure 2 • Graphe de successions des actes de langage dans SPLACH**

L'outil de conversation de SPLACH est présenté sur la figure 3. Il se présente sous la forme d'une arborescence d'interventions. Le principe est de lier chaque message à celui auquel il répond ou réagit. Les nouveaux sujets de discussion sont placés à la racine de l'arbre, les autres se raccrochant aux messages existants. L'avantage de cette représentation est de tenir compte des fils de discussion et donc des sujets de conversation. Sur la figure 3, nous pouvons voir un fil de discussion contenant six interventions. Chaque intervention commence par une icône symbolisant l'acte de langage sélectionné par l'utilisateur. Pour intervenir, l'utilisateur clique soit sur un message existant, soit à la fin de la conversation sur la ligne « cliquer ici pour commencer une nouvelle discussion ». Dans les deux cas un menu apparaît près de la souris et propose la liste des actes possibles (d'après les enchaînements spécifiés ci-dessus). L'utilisateur sélectionne alors l'acte souhaité puis saisit le texte de son intervention de façon libre. Le message est alors inséré à la bonne place dans l'arborescence. Dans l'exemple ci-dessous, un clic sur une question de Michel a fait apparaître un menu déroulant permettant de « répondre » ou de « questionner ».

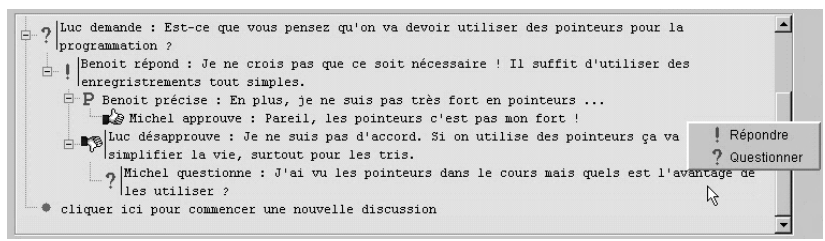

**Figure 3 • L'outil de conversation textuelle synchrone structurée**

L'un des objectifs de cet outil est de pouvoir l'utiliser pour l'analyse automatique des conversations. Dans le cas de SPLACH, un agent informatique recueille les données afin de déterminer automatiquement quatre profils types de comportement sociaux : animateur, vérificateur, quêteur et indépendant (Pléty, 1996). Des formules heuristiques permettent de calculer un profil pour chaque utilisateur en fonction de sa participation et des actes langages qu'il utilise. Cette recherche et en particulier des résultats d'expérimentation de l'outil sont détaillés dans (George, 2001).

Cet outil et l'utilisation qui en est faite posent cependant quelques questions. En effet, l'outil informatique support des discussions n'est pas neutre dans l'activité des utilisateurs et de ce fait cela entraîne un biais dans le calcul des profils comportementaux. De ce fait, l'analyse qui est faite ici est celle d'une conversation médiatisée par un outil. Le calcul est effectué par rapport à l'activité de conversation dans un contexte précis et ne peut prétendre refléter les comportements des utilisateurs d'une façon générale.

### 3.5. Les atouts pédagogiques de la structuration des communications médiatisées synchrones

En reprenant les quatre axes définis dans la partie 2, nous pouvons classer l'outil C-Chene dans les outils visant à orienter vers des formes d'interactions particulières. En effet, l'objectif premier est de faciliter la gestion de l'interaction en orientant la discussion à l'aide d'ouvreurs de phrases spécifiques à l'activité. Le chat de l'environnement SPLACH se classe quant à lui dans les outils proposant une présentation particulière des fils de discussion (en l'occurrence un chat avec une structuration arborescente) ainsi que dans les outils cherchant à faciliter l'analyse automatique des discussions (utilisation d'ouvreurs de phrase qui sont analysés pour déterminer des profils de comportement).

Nous pouvons regretter que ce type d'outil reste à l'état expérimental. Les causes sont certainement multiples mais la principale vient du fait que ces outils sont conçus dans le cadre d'une recherche ciblée ayant un objectif défini. Une fois la recherche terminée, l'outil n'est alors plus utilisé. Une exception existe cependant car le chat structuré de l'environnement SPLACH a été repris dans un outil autonome nommé Oscar (Delium, 2003). Dans cet outil, les actes de langage et les grammaires d'enchaînement sont entièrement paramétrables. De plus, Oscar permet à la fois les discussions synchrones et asynchrones. L'outil téléchargeable (Oscar, 2006) est régulièrement utilisé.

Les deux exemples présentés montrent les possibilités offertes par la conception d'outils spécifiques structurant les discussions entre apprenants. Nous pouvons retenir qu'une structuration de la conversation par une interface peut être considérée comme utile dans un contexte d'apprentissage pour les raisons suivantes :

– le fait de typer un message amène l'utilisateur à se demander ce qu'il veut dire et a donc une valeur éducative (Winograd, 1987 ; Flores *et al.*, 1988) ;

– une interface structurée encourage les utilisateurs à se centrer davantage sur la tâche (Baker et Lund, 1997 ; Jermann et Schneider, 1997) ;

– l'utilisation d'ouvreurs permet de concevoir des systèmes d'analyse automatique de discussions (McManus et Aiken, 1995 ; George, 2001).

Pour que les actes de langage soient facilement utilisables, il ne faut pas en définir un trop grand nombre. Si ce nombre d'actes est raisonnable (une douzaine), ils sont souvent choisis à bon escient par les utilisateurs. Par exemple, dans une expérimentation de SPLACH (George, 2001), seuls

10 à 15% des actes sélectionnés ne correspondent pas au contenu du message. Par ailleurs, la structuration avec des actes doit être ergonomique. Les contraintes imposées sont cependant inévitables mais doivent être contrebalancées par les atouts engendrés. Par exemple, les effets positifs attendus peuvent être de favoriser certaines formes d'interactions entre étudiants, interactions pouvant conduire à des apprentissages tels qu'apprendre à argumenter ou développer la réflexion.

Dans cette partie, nous nous sommes uniquement intéressés à des modèles et outils de communication pour des échanges synchrones. Certaines caractéristiques sont bien spécifiques à la modalité synchrone, comme le fait d'avoir des messages assez courts (le plus souvent, envoi d'une seule phrase) qui se prêtent bien à la structuration par des actes de langage. Pour des communications médiatisées asynchrones, les modèles et structurations ne peuvent être les mêmes car il y a rarement unicité d'un acte de langage pour un message. Nous explorons dans la prochaine partie des recherches sur la conception, pour un contexte éducatif, d'outils de communication asynchrone.

## 4. Conception d'outils de communication asynchrone : les forums structurés

Dans cette partie, nous nous focalisons sur les forums de discussions et plus particulièrement dans un contexte de formation en ligne. Les forums sont des outils qui supportent les communications écrites asynchrones. Cette possibilité de communication asynchrone est nécessaire en formation en ligne car bien souvent la distance entre les personnes n'est pas seulement physique mais également temporelle (fuseaux horaires ou moments de connexions différents).

### 4.1. La surestimation du potentiel éducatif des forums

Toutes les plates-formes de formation en ligne intègrent des outils de type forum. Ces outils ne sont pas spécifiques à l'éducation mais des bénéfices pédagogiques sont attendus comme, par exemple, stimuler la participation active des apprenants, augmenter leur motivation, créer le sentiment d'appartenance à un groupe, permettre le monitorat entre pairs. Dans les domaines du *Computer Supported Collaborative Learning* (CSCL) et des sciences de l'éducation, de nombreux travaux ont tenté de démontrer le potentiel de ces outils de communication comme support à l'apprentissage collectif. Nous n'avons qu'à citer le symposium Symfonic

(Symfonic, 2005) qui s'est tenu à Amiens sur le thème des forums en éducation. Ce symposium a eu pour objectif de faire un point sur les modes d'étude et d'analyse des forums utilisés dans des contextes d'apprentissage. Les expériences relatées montrent que l'outil forum à lui seul ne peut être le déclencheur d'une dynamique collective. La mise en place d'une situation pédagogique favorisant les interactions est bien souvent nécessaire. Ainsi, lors d'activités pédagogiques individuelles, le simple fait de fournir des outils classiques de communication à distance n'est pas suffisant pour faire émerger des interactions entre apprenants et ainsi favoriser la construction collective de connaissances (Gommer et Visser, 2001 ; Hotte et Pierre, 2002). Les interactions qui émergent de façon naturelle en présence n'apparaissent pas de manière aussi spontanée à distance.

Nous présentons dans les sous-parties suivantes deux approches de conception d'outils de forum, Mailgroup et CONFOR, qui ont pour objectif de pallier certains problèmes récurrents et notamment la difficulté à mettre en évidence la dynamique temporelle des échanges et la difficulté à pouvoir discuter de façon contextuelle. En reprenant la classification de la partie 2, les deux outils visent à « faciliter le déroulement des interactions » et l'outil CONFOR (partie 4.3) cherche de plus à « inciter à la création d'interactions ».

### 4.2. Mailgroup : un forum temporellement structuré

La recherche sur les interactions sociales tient une place importante dans les études concernant l'utilisation des forums par les communautés d'apprentissage. En particulier, en observant de telles interactions, (Reyes et Tchounikine, 2004) ont fait ressortir certains comportements temporels d'étudiants utilisant les forums : dans un laps de temps relativement court, les étudiants répondent généralement à plusieurs messages situés dans des fils de discussion différents. Ainsi, une analyse de différents newsgroups a fait ressortir qu'un quart des messages sont envoyés de manière consécutives (dans différents fils de discussion). Partant de ce constat, ces chercheurs ont conçu un outil permettant de faire ressortir ce phénomène dans les forums.

L'outil proposé se nomme « Mailgroup ». Il a la particularité de faire ressortir visuellement les messages envoyés par un étudiant lors d'une même période de connexion. Reyes (Reyes, 2005) définit ainsi la notion de « session » qui correspond à un groupe de messages envoyés consécutivement par une même personne. La figure 4 montre l'interface du forum

Mailgroup. Sur la partie supérieure, un graphe permet aux utilisateurs de voir la structure des fils de discussion qui évoluent linéairement et en parallèle les uns par rapport aux autres. Si un utilisateur contribue dans une session à plusieurs fils de discussion, ses messages sont placés dans une même colonne. De cette façon, les utilisateurs peuvent percevoir l'ordonnancement temporel et les fils de discussion avec une seule et même visualisation. Avec cette représentation, les utilisateurs sont davantage conscients de la structure des interactions de leur communauté.

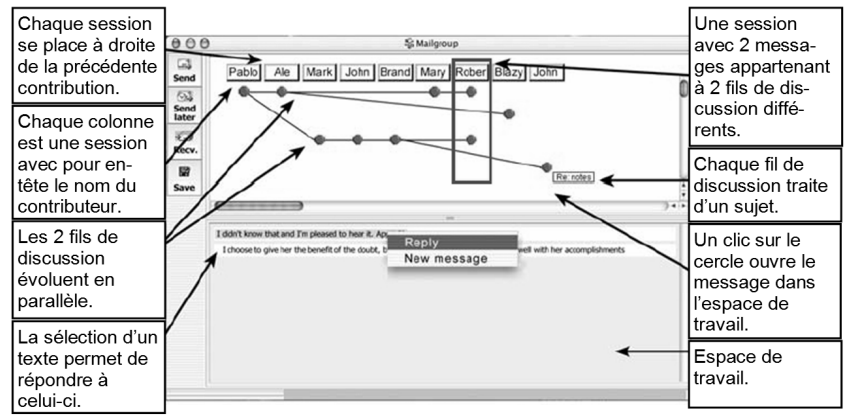

**Figure 4 • L'interface du forum « Mailgroup » (Reyes, 2005)**

L'objectif de l'outil Mailgroup est d'améliorer et de faciliter l'émergence d'interactions d'apprentissage dans les forums de discussion. En effet, la mise en évidence des tours de parole doit pouvoir contribuer à une meilleure gestion de ceux-ci. L'objectif final est de dépasser l'échange d'information entre les participants et de provoquer des connexions entre des idées qui n'avaient pas de lien *a priori* (Bellamy et Woolsey, 1998). Dans Mailgroup, la structuration en sessions va dans ce sens en regroupant des messages qui auraient été dispersés dans différents fils de discussion avec des outils de forum classique. La notion de session tente également de mettre en évidence les tours de paroles pour une meilleure coordination des participants.

Une expérimentation du prototype Mailgroup s'est déroulée avec 15 personnes (Reyes et Tchounikine, 2004). Les participants étaient des enseignants qui réalisaient une activité collaborative à distance durant un mois et demi dans le cadre d'une formation sur les technologies de l'information et de la communication (TIC). La tâche concernait l'analyse collaborative d'intégration des TIC dans le milieu éducatif. L'objectif de l'expérimentation était d'avoir un retour de la part des utilisateurs. Les contributions sur le forum ont été étudiées afin d'examiner l'impact de la visualisation sous forme de sessions. Par ailleurs, les impressions des utilisateurs ont été recueillies grâce à des questionnaires. L'expérience montre tout d'abord que le logiciel n'a pas changé le comportement temporel des utilisateurs : un quart des messages sont écrits de manière consécutive. L'expérience ne fournit pas de résultat précis sur l'aide qu'apporte la structuration des messages en session. Néanmoins, les questionnaires indiquent que trois quarts des 15 participants considèrent que la visualisation et l'organisation temporelle des messages permettent de mieux suivre le développement des conversations.

Une des limites que l'on peut trouver à ce système est que deux messages d'une même session ne sont pas forcément liés par leur contenu. MailGroup met ainsi davantage en avant les participants plutôt que les sujets de discussion. D'autres tests seraient nécessaires pour montrer l'impact positif de l'outil sur la gestion des discussions.

### 4.3. CONFOR : un forum contextuel

Les forums utilisés dans un contexte pédagogique sont dans la plupart du temps utilisés comme outil de communication des apprenants avec leurs enseignants. Les relations verticales apprenant-enseignant sont nettement prédominantes dans les formations en ligne (Hotte, 1998). Une des causes principales provient des outils mêmes et plus particulièrement de leurs faibles liens avec les activités d'apprentissage (Looi, 2001). Une des conséquences est que cette séparation n'incite pas à utiliser ces outils de discussion : « *The problem with content-related communication often is, that it doesn't occur because it is a separate activity that is not integrated in the course* » (Gommer et Visser, 2001). Le projet CONFOR (*CONtextual FORum*) vise à déterminer comment rapprocher les activités de discussion des activités d'apprentissage avec des outils de communication adaptés (George, 2003).

Le modèle de forum CONFOR proposé est fondé sur deux caractéristiques :
– une contextualisation du forum qui se traduit par un affichage automatique de la partie du forum qui contient les messages pertinents par rapport à l'activité de l'apprenant ;
– une structuration du forum qui permet d'organiser les messages entre eux mais aussi par rapport aux contenus et aux activités. Deux structu-

rations possibles du forum ont été étudiées, celles-ci étant complémentaires :

- selon la structuration pédagogique d'un cours en ligne (George, 2003). Typiquement, la structure du forum reprend par exemple le scénario pédagogique d'un cours divisés en parties et sous-parties.
- selon des connaissances abordées par un objet pédagogique (George, 2004). Les messages du forum sont dans ce cas reliés à des éléments de connaissances organisés entre eux (structure conceptuel d'un cours).

La version actuelle du logiciel CONFOR articule les deux structurations décrites ci-dessus (Figure 5). Ainsi, quand un étudiant consulte un objet pédagogique, le forum contextuel lui présente (1) les messages en lien avec son activité et (2) les messages en lien avec les connaissances abordées par la ressource. L'apprenant est ainsi à même de discuter sur l'activité (pour parler de l'organisation du cours par exemple) ou sur le contenu (pour discuter du fond). Bien sûr, ce type de forum n'exclut pas l'utilisation d'autres forums plus généraux qui permettent par exemple la socialisation ou la résolution de problèmes techniques.

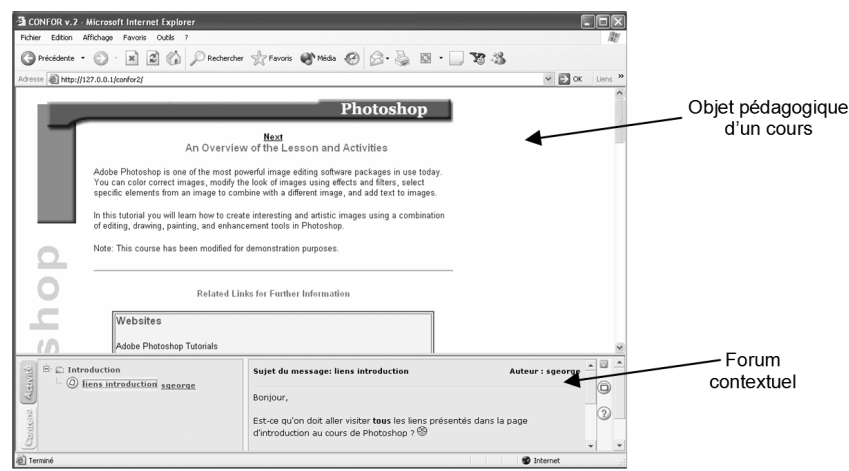

**Figure 5 • L'interface du forum CONFOR**

Sur la figure 5, la partie supérieure de la fenêtre représente un objet pédagogique d'un cours en ligne. En bas de ce cours se trouve l'affichage contextuel du forum. Ainsi, sur l'écran ci-dessus, l'utilisateur réalise une activité de son cours et il voit en même temps les messages du forum correspondant à cette activité (onglet « activité » du forum). En cliquant sur l'onglet « contenu » du forum (en bas à gauche), il voit alors les discussions liées aux connaissances abordées dans l'objet pédagogique. Dans les deux cas, la vue contextuelle du forum représente une partie d'un forum global. D'un point de vue utilisation, le forum contextuel peut être redimensionné par l'utilisateur ou bien mis dans une fenêtre flottante toujours visible.

D'un point de vue technique, l'outil s'appuie sur les normes actuelles. Ainsi, la spécification SCORM (Dodds et Thropp, 2004) est utilisée pour détecter l'ouverture des objets pédagogiques. Par ailleurs, les metadonnées de type LOM (*Learning Object Metadata*) sont utilisées pour retrouver les éléments de connaissances abordés par un objet pédagogique. L'utilisation de ces normes a entre autres pour avantage de rendre l'outil CONFOR facilement interopérable avec toute plate-forme qui respecte celles-ci.

Une expérimentation de CONFOR s'est déroulée sur 8 mois auprès de 70 étudiants encadrés par 2 tuteurs à la Télé-université du Québec. Le but de l'évaluation était d'étudier l'utilisabilité et l'utilité de CONFOR en ayant recours à des questionnaires, des entrevues et des observations régulières du forum couplés avec l'analyse de traces informatiques. D'un point de vue quantitatif, les traces montrent une utilisation importante de la vue contextuelle du forum par rapport à une vue globale qui était aussi proposée. Ainsi, 4.5 fois plus de messages ont été ouverts en mode contextuel qu'en mode global. Par ailleurs, la vue contextuelle semble favoriser l'envoi de messages (7.5 fois plus de messages envoyés en mode contextuel qu'en mode global). D'un point de vue qualitatif, les réponses aux questionnaires indiquent que les étudiants ont globalement apprécié le forum contextuel. Ils pensent que cet outil incite vraiment la lecture de messages postés par d'autres étudiants. Il ressort également que CONFOR aide les étudiants à mieux cibler les messages pertinents pour leurs activités.

Une nouvelle version de CONFOR est actuellement en cours de développement dans le cadre d'un projet franco-québecois entre l'INSA de Lyon et la télé-université du Québec - UQAM. L'idée est de proposer CONFOR comme un *web service* pouvant être appelé par les plates-formes d'*e-learning*. Un des objectifs est de pouvoir l'expérimenter à plus grande échelle.

### 4.4. Les bénéfices de la structuration des fils de discussion

Les deux outils de forums présentés dans cette partie ont la particularité de structurer et de présenter les fils de discussion selon différents points de vue : le moment de l'écriture, le contexte du message ou bien encore les connaissances abordées. L'utilisation de présentations alternatives des discussions a clairement pour objectif de *faciliter le déroulement des interactions*. En effet, les structurations habituelles des forums, sous forme chronologique (en tableau) ou hiérarchique (en arbre) (Dimitracopoulou et Bruillard, 2007), ne permettent pas aisément de relier entre eux des messages qui sont « proches » d'un point de vue contenu ou écrit dans une même période de temps (donc proches du point de vue de l'auteur). Pour l'outil CONFOR, un deuxième objectif est mis en avant : celui d'*inciter la création d'interactions*. En effet, la présentation contextuelle des messages permet d'afficher automatiquement les messages pertinents par rapport à l'activité de l'utilisateur. En affichant ces messages, l'utilisateur est incité à les lire et à réagir.

Au-delà des bénéfices avancés par les concepteurs, nous pouvons cependant nous demander si le morcellement des discussions ne peut en contrepartie être source de problèmes. En effet, la structuration oblige à découper les échanges pour les relier à des références (une référence au temps, une référence à un contenu, etc.). La question qui se pose alors est de savoir si le fait de morceler les discussions n'enlève pas la vision globale des échanges, vision qui est souvent nécessaire pour comprendre le sens général d'un ensemble d'idées émises. Le fait de faire ressortir certaines dépendances va par la même occasion en supprimer d'autres. Comme bien souvent en EIAH, c'est la question de la granularité qui se pose ici. Jusqu'où devons-nous structurer les discussions ? Plus la structuration est fine et plus les échanges seront morcelés et inversement aucune structuration rendra un outil difficilement exploitable. Il est bien évidemment impossible d'apporter une réponse générale sur la bonne granularité à choisir. Le contexte de la formation, l'activité à réaliser, le type de public, le nombre d'utilisateurs sont autant de paramètres à prendre en compte pour choisir le niveau de structuration et le retour d'usage sera bien souvent le seul moyen de vérifier la justesse du choix. Par ailleurs, une même structuration ne convient pas forcément aux deux modalités de lecture et d'écriture dans un forum. Il faut donc bien analyser l'utilisation réelle des utilisateurs et ne pas hésiter à proposer des représentations multiples d'un même forum.

### 5. Conception de systèmes de mise en relation interpersonnelle

Les différents outils décrits précédemment, aussi bien synchrones qu'asynchrones, vont dans le sens d'améliorer la qualité des échanges entre apprenants et de donner à des conversations une dimension pédagogique. Cependant, parfois, la problématique se situe en amont : les utilisateurs n'ont pas forcément l'idée de communiquer, voire la motivation nécessaire. Nous décrivons dans cette partie un outil particulier, qui vient en support à la formation proprement dite, dédié au développement de réseaux d'entraide. L'idée principale est de faciliter les rapprochements entre apprenants dans leur recherche d'aide pour une situation particulière, par exemple pour la manipulation des outils. L'objectif est donc ici d'inciter et d'assister la mise en relation entre apprenants.

#### 5.1. Les réseaux d'entraide pour briser l'isolement

Les observations réalisées au sein de plusieurs environnements de formations à distance (formation continue, campus numérique, formation intra-entreprise) ont mis en évidence trois formes de collectif : l'anomie qui est la conséquence d'une insuffisance d'interaction (Cusson, 1992), le réseau qui permet la transmission de ressources matérielles ou informationnelles portées par des ressources relationnelles (Lemieux, 1999) et la communauté dont les membres partagent un « ensemble d'évidences et de pratiques culturelles communes » (Zarifian, 1996). Il a été montré que l'apparition de réseaux d'entraide et de formes communautaires vient consolider le processus d'apprentissage dans des contextes de formation universitaire et de formation professionnelle (Foucault *et al.*, 2002). Ainsi, l'analyse des interactions entre apprenants (via les traces laissées sur les forums de discussion) confirme l'existence de pratiques d'échanges qui relèvent de l'entraide, pour compenser le sentiment d'isolement et le manque de réactivité et/ou de disponibilité des tuteurs (Foucault *et al.*, 2003a ; Foucault *et al.*, 2003b). L'analyse de ces traces peut également mettre en évidence différents types de relations de type communautaire qu'il est possible de corréler aux différentes situations d'apprentissage (Haythornthwaite, 1998).

Aujourd'hui, la constitution de réseaux d'entraide ou de communauté demeure une initiative informelle provenant des apprenants et qui semble encore insuffisamment prise en compte par les dispositifs de e-formation. Dans un contexte plus général, des sites comme *Friendster*, *LinkedIn*, *Flickr*, etc. connaissent un vif succès et permettent aux internautes de développer des liens sociaux. Dans le monde de l'éducation et de la formation en ligne, nous pouvons citer *elgg.net* qui propose un *ePortfolio* social où l'apprenant peut détailler son CV en ligne et se constituer des communautés d'amis ou de contacts.

L'initiave Formatis décrite dans la partie 5.2 montre l'opportunité de créer des réseaux d'entraide autour de la formation à proprement parler, et montre que la constitution de réseaux sociaux peut aider à raviver la motivation des apprenants, à stimuler leur participation, à leur faire prendre conscience de la composante sociale inhérente à l'apprentissage et ainsi favoriser l'enrichissement mutuel. Formatis n'a pas pour objectif de supporter une communauté durable d'apprenants dans le sens défini ci-dessus. Le but est de favoriser l'émergence de regroupements d'apprenants de façon moins structurée, ce qui rentre dans la définition des réseaux d'entraide. Il est évident que ces réseaux d'entraide seuls n'ont pas de sens, et il convient de bien les intégrer à des environnements pédagogiques offrant des outils de communication comme ceux décrits dans les parties précédentes.

### 5.2. Formatis : faciliter l'émergence de réseaux d'entraide

Les réseaux d'entraide sont le reflet d'interactions entre apprenants suscitées par un besoin de soutien. Ils se matérialisent par des échanges de conseils, sur et autour de la formation (compréhension d'un cours, ou encore soutien pour l'utilisation des outils informatiques) mais aussi, plus simplement, des discussions (un exemple qui revient souvent dans les résultats d'enquête cités en introduction est la possibilité d'interagir autour des questions d'organisation de son travail à distance).

Le prototype Formatis, basé sur la mise en relation interpersonnelle, a pour but de réifier ces réseaux d'entraide, d'en faciliter le développement et de créer ainsi une dynamique d'échanges visant à mobiliser les apprenants. Il s'agit en quelque sorte de recréer certaines caractéristiques d'une classe en présentiel. L'un des points essentiels pour l'apprenant est avant tout de pouvoir se situer par rapport aux autres apprenants. Formatis propose ainsi la visualisation graphique de son avancée dans le parcours de formation, mais aussi la possibilité de consulter l'identité des autres ou encore de rechercher les apprenants présentant des profils similaires.

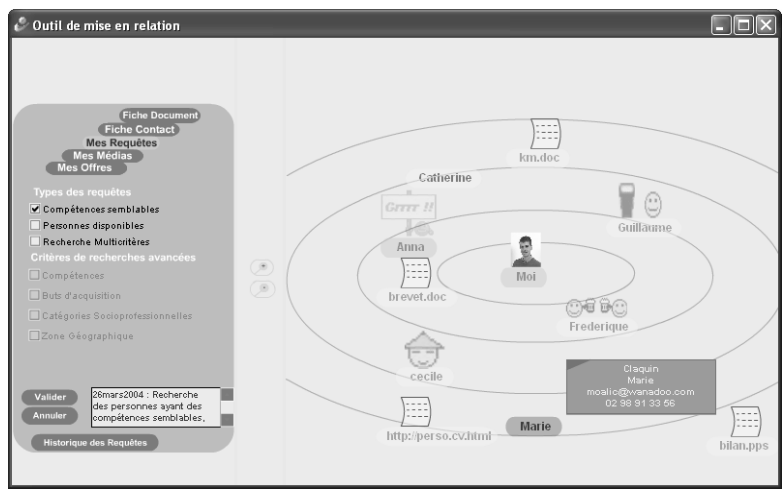

**Figure 6 : Visualiser ses pairs avec Formatis**

Sur la figure 6, nous voyons la fenêtre principale de l'outil Formatis. La partie gauche permet de décrire une requête ou l'objet sélectionné, et la partie graphique à droite représente « l'espace » de la formation, avec les pairs répondant à une requête. Dans notre exemple, l'utilisateur recherche des personnes ayant les mêmes compétences que lui. Des codes de couleurs donnent des indications (la figure en couleur peut être consultée sur le site www.sticef.org). Ainsi, les noms sur fond orange représentent les personnes connectées (« moi » et Anna sur la figure 6). Les personnes en écriture rose sur fond jaune sont des contacts personnels de l'utilisateur mais ils ne sont pas connectés (Cécile, Frédérique, Guillaume). Les personnes en écriture noire sont celles qui peuvent être intéressantes vis à vis de la requête mais qui ne sont pas dans la liste des contacts de l'utilisateur (Catherine, Marie). Marie est sur fond mauve car la souris est placée au dessus d'elle, ce qui fait qu'une note apparaît contenant ses informations principales. Par ailleurs, des documents peuvent aussi être pertinents vis-à-vis de la requête et ils sont alors indiqués en rouge sur fond jaune (« km.doc », « bilan.pps », …).

Dans une logique gagnant-gagnant, et afin d'impliquer chacun, un onglet « Mes offres » permet de définir en quoi la personne utilisatrice peut aider ses pairs (Figure 7). Au-delà du choix qui est fait, cet onglet permet une prise de conscience de l'appartenance au groupe et du rôle potentiel que l'on peut y jouer. L'onglet « Fiche contact » permet de voir le détail du profil d'une personne. Une autre version permet de voir le positionnement dans les formations, les compétences et diplômes.

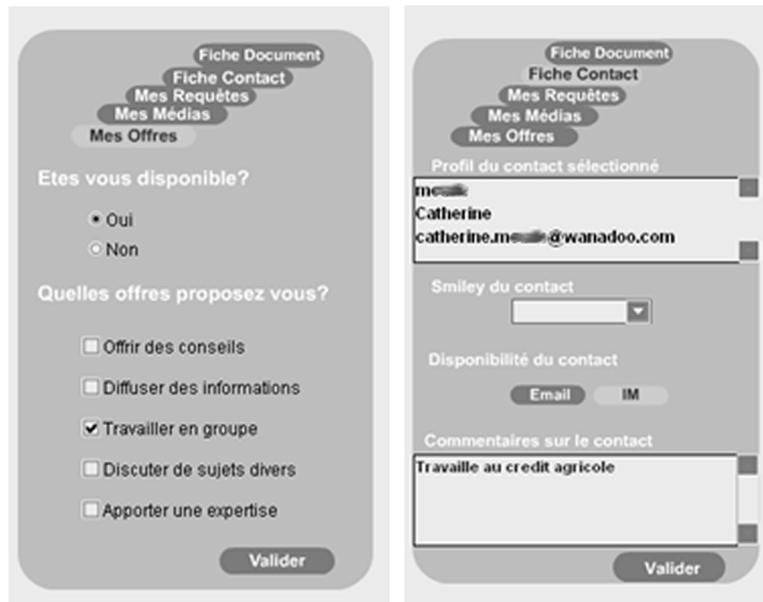

**Figure 7 : Se faire aider, mais aussi aider les autres**

Formatis a vocation à être proposé comme un outil annexe d'un dispositif d'apprentissage, accessible depuis d'autres outils dédiés à la formation, tels que des forums). Dans cette perspective, Formatis se veut potentiellement utilisable par différentes plates-formes e-learning et interopérable avec d'autres outils, tels que des forums. L'interopérabilité est assurée par l'utilisation de technologies telles que *Jabber* pour les interactions entre le module de mise en relation et les outils de communication, ou encore SOAP pour les échanges de données entre le module de visualisation et la base de données. L'interface est en Flash (Macromedia) ce qui la rend compatible avec les navigateurs web du moment.

L'outil Formatis n'a par encore été testé sur des populations réelles d'apprenants. Toutefois, il a été décliné au monde professionnel dans une logique de socialisation du *Knowledge Management* (Bothorel *et al.*, 2005). Cette version de l'outil est portée par une équipe projet de grande taille (50 collaborateurs avec des compétences différentes) dans une société de R&D et a pour vocation de favoriser l'établissement de réseaux d'entraide, voire de communautés de pratique. Dans cette version l'analyse de traces a été mise en œuvre (logs de manipulation de documents) et l'automatisation d'un processus basé sur les graphes générant des profils d'intérêts présente graphiquement les proximités entre collaborateurs selon une thématique donnée. Les perspectives à moyen terme sont d'introduire des algorithmes du domaine des réseaux sociaux pour qualifier au mieux les positions dans le graphe et faire émerger des communautés de pratique.

### 5.3. Assister la mise en relation inter-apprenants

Nous avons décrit dans cette partie un système permettant de rechercher les personnes les plus à mêmes d'aider ou de conseiller un apprenant pour une situation particulière. Ce type de système s'inscrit complètement dans la catégorie des outils *incitant la création d'interactions*. La difficulté première dans cette catégorie d'outil est de trouver des algorithmes fournissant des résultats pertinents lors d'une requête. Pour une description plus détaillée des différents algorithmes possibles, le lecteur pourra se référer par exemple à (Becks *et al.*, 2003). Le système Formatis présenté n'intervient pas directement dans le processus d'apprentissage, mais apparaît comme un support intéressant pour tout dispositif de formation en ligne de par la facilitation de création de liens sociaux. Ce point devient d'autant plus pertinent que la formation en ligne s'intéresse à présent aux situations de mobilité.

## 6. Conclusion et perspectives

Au travers des différents exemples donnés dans cet article, nous avons cherché à montrer que des outils de communication médiatisée pouvaient être spécifiquement conçus dans le but de produire des effets positifs dans un contexte éducatif. Les objectifs recherchés peuvent être divers et nous avons identifié quatre principaux axes : faciliter les interactions, orienter les interactions, inciter les interactions et enfin favoriser l'analyse automatique des interactions. Dans tous les cas, ce type d'outils impose des contraintes particulières pour l'utilisateur mais apportent en même temps des facilités dans la structuration des échanges, voire des impacts bénéfiques sur l'apprentissage, même si ceux-ci restent bien souvent difficilement mesurables. En effet, identifier de tels impacts n'est pas trivial. Seule une utilisation des systèmes dans des situations réelles et à une échelle importante pourrait donner des résultats précis à ce sujet. Pour le moment, les expérimentations menées se limitent souvent à observer l'impact sur la communication (augmentation du nombre de messages, aspect qualitatif des discussions,…) qui, sans donner d'indications précises sur l'amélioration de l'apprentissage, apporte néanmoins des informations sur l'amélioration des conditions pour l'échange entre pairs.

Nous pouvons dégager des perspectives de cette étude. Afin de faciliter la régulation des discussions, il faudrait pouvoir prendre en compte non seulement les productions résultant des outils de communication (par exemple les échanges dans un chat, les fils de discussion d'un forum, …) mais aussi des informations sur les processus liés qui ne laissent pas d'information dans la discussion elle-même (par exemple l'ouverture d'un message par un utilisateur ou bien encore le parcours dans un forum arborescent). Cette préoccupation rejoint la question des « traces » qui est d'actualité dans le domaine des EIAH. La capacité à produire des traces pertinentes et exploitables des activités de communication peut être utile aux différents acteurs de la formation en ligne. Par exemple, les apprenants pourraient disposer d'informations synthétiques sur le comportement des autres (notion d'*awareness*) et sur leur propre comportement. Pour l'enseignant, l'enjeu est de l'aider dans son activité de suivi de situations d'apprentissage collective à distance. Enfin, pour le chercheur, il s'agit d'améliorer les observations de l'usage des outils. Selon nous, il faudrait bien distinguer deux types de traces : les traces comme produits de l'activité et les traces des actions effectuées sur les outils d'interaction. Les premières peuvent conduire à l'analyse automatique des contenus alors que les deuxièmes peuvent être utiles pour analyser des comportements. Dans les deux cas, cela nécessite de définir des indicateurs pertinents. Pour une étude plus précise à ce sujet, le lecteur pourra se référer à l'article de Dimitracopoulou et Bruillard (2007) dans ce numéro.

Pour conclure, nous pouvons émettre le souhait que se développe une meilleure articulation des outils de communication entre eux. Sans aller vers une normalisation trop stricte, des spécifications et des formats d'échange de données pourraient aider à utiliser conjointement des systèmes qui répondent à des besoins particuliers. Par ailleurs, une plus grande adaptabilité des systèmes (facilité de paramétrage, malléabilité) serait également un avantage pour une meilleure adoption des outils par les systèmes éducatifs, ce qui se traduirait par des utilisations à plus grande échelle.